\documentclass[%
 aip,
 jmp,%
 amsmath,amssymb,
 reprint,%
]{revtex4-1}

\usepackage{graphicx}
\graphicspath{{Figures/}}
\usepackage{dcolumn}
\usepackage{bm}
\usepackage{subfigure}

\usepackage{algorithm}
\usepackage{algpseudocode}
\usepackage{amsmath,amssymb,amsfonts}
\usepackage{graphics}
\usepackage{color}

\begin{document}


\title[]{Connecting implicit and explicit large eddy simulations of two-dimensional turbulence through machine learning}

\author{R. Maulik}
\author{O. San}%
\email{osan@okstate.edu}
\author{J. D. Jacob}%

\affiliation{ 
School of Mechanical \& Aerospace Engineering, Oklahoma State University, Stillwater, Oklahoma - 74078, USA.
}%

\date{\today}

\begin{abstract}
In this article, we utilize machine learning to dynamically determine if a point on the computational grid requires implicit numerical dissipation for large eddy simulation (LES). The decision making process is learnt through \emph{a priori} training on quantities derived from direct numerical simulation (DNS) data. In particular, we compute eddy-viscosities obtained through the coarse graining of DNS quantities and utilize their distribution to categorize areas that require dissipation. If our learning determines that closure is necessary, an upwinded scheme is utilized for computing the non-linear Jacobian. In contrast, if it is determined that closure is unnecessary, a symmetric and second-order accurate energy and enstrophy preserving Arakawa scheme is utilized instead. This results in a closure framework that precludes the specification of any model-form for the small scale contributions of turbulence but deploys an appropriate numerical dissipation from explicit closure driven hypotheses. This methodology is deployed for the Kraichnan turbulence test-case and assessed through various statistical quantities such as angle-averaged kinetic energy spectra and vorticity structure functions.  Our framework thus establishes a direct link between the use of explicit LES ideologies for closure and numerical scheme-based modeling of turbulence leading to improved statistical fidelity of \emph{a posteriori} simulations. 
\end{abstract}

\maketitle


\section{Introduction}


Over the past decade, advances in data collection and increasing access to computational resources have led to a revolution in the use of data-driven techniques for the solution of complex inverse problems. One such problem is that of turbulence, the multiscale nature of which causes extreme computational demands for most practical systems. As a result, turbulence requires the use multiple modeling approximations for the higher wavenumbers which remain unsupported by computational degrees of freedom. One such modeling approach is that of large eddy simulation (LES) \cite{sagaut2006large}, which attempts to simulate the evolution of the smaller wavenumbers while the unresolved frequencies are modeled by an algebraic or differential equation. As such, the basic premise of LES is extendable to many partial differential equation systems with quadratic non-linearities. The procedure of modeling these smaller scales is often denoted \emph{closure} due to insufficient knowledge about higher-order wavenumber interactions with the coarse-grained system \cite{berselli2006mathematics} and remains vital for the accurate computation of many applications \cite{hickel2014subgrid,yu2016dynamic,zhou2018structural}. From an LES point of view, the closure problem may be considered to be dominated by commutative errors in the calculation of the non-linear term as well as the defects associated with commutative errors stemming from the dynamic term. In this study, we focus on the former.

There are two main schools of thought when it comes to the LES of the Navier-Stokes equations. The first of these promotes the use of explicit closures. Explicit LES argues for the utilization of closures in the form of sub-grid models specified as algebraic or differential equations for the unresolved scales. These are built on intuitive reasoning of how the losses of coarse graining the Navier-Stokes equations may be incorporated into an LES deployment. Some of the most notable sub-grid closure strategies are those given by the eddy-viscosity hypothesis. Within the context of the Navier-Stokes equations, it is generally accepted that the finer scales are dissipative at the Kolmogorov length scales \cite{kolmogorov1941local} and therefore, most turbulence models seek to specify a sub-grid dissipation \cite{frisch1995turbulence}. Most sub-grid models can be traced back to the seminal work of Smagorinsky \cite{smagorinsky1963general}, where a model was proposed based on the concepts of an effective eddy-viscosity determined by an \emph{a priori} specified mixing length and a $k^{-5/3}$ scaling recovery for the kinetic energy content in the wavenumber domain. Similar hypotheses have also been used for two-dimensional turbulence \cite{leith1968diffusion} (often utilized as a test-bed for geophysical scenarios, for instance see works by Pearson \textit{et al.}\cite{pearson2018log,pearson2017evaluation}), for approximating the $k^{-3}$ cascade in two-dimensional turbulence and generally have their roots in dimensional analysis related to the cascade of enstrophy. These models may also be classified as \emph{functional} due to the phenomenological nature of their deployment and represent the bulk of explicit LES turbulence models used in practical deployments. Explicit LES closures may also be specified through the specification of a low-pass spatial filter to account for the unresolved scales \cite{bardina1980improved,stolz1999approximate,layton2003simple,mathew2003explicit} where phenomenology is bypassed but ansatz are provided for the bulk dissipative nature of the smaller scales through the control of a characteristic filter-width. In either scenario, (i.e., whether structural or functional), the choice of the phenomenology (or dissipation control parameter) plays a key role in the successful calculation of accurate \emph{a posteriori} statistics. In contrast, the implicit LES (or ILES) approach utilizes numerical dissipation to model the unresolved scales in a turbulent flow \cite{grinstein2007implicit,el2017investigation,Margolin2018}. In essence, the predominantly dissipative effects of the smallest scales are replicated through an artificial numerical dissipation via a biased discretization used in the calculation of the non-linear advective term \cite{thornber2007implicit,debonis2013solutions}. The ILES approach is popular due to reduced algorithmic complexity and represents a union of turbulence modeling and shock capturing mechanisms but is often criticized due to the difficulties involved in quantifying the correct amount of dissipation in a turbulent flow evolution. This results in ILES methods often proving robust and stable but overly dissipative. In this work, we propose a machine learning algorithm to enable selective dissipation within an ILES deployment through the use of explicit LES concepts during the training of the learning framework. 



The past few years have seen a rapid increase in the use of data-driven techniques for the spatio-temporal modeling of dynamical systems \cite{schmidt2009distilling,bright2013compressive,xiao2015non,ma2015using,gautier2015closed,brunton2016discovering,schaeffer2017learning,raissi2017machine,mohan2018deep,raissi2018hidden,rudy2018deep,san2018neural,wan2018data,kim2018deep,muravleva2018application,jin2018prediction}. When it comes to turbulence, some widely used strategies for inference include symbolic regression \cite{weatheritt2016novel,weatheritt2017development,weatheritt2017hybrid}, where functional model-forms for Reynolds-averaged Navier-Stokes (RANS) deployments were generated through evolutionary optimization against high-fidelity data. Other techniques incorporating Bayesian ideologies have also been used, for instance by Xiao \textit{et al.}\cite{xiao2016quantifying} where an iterative ensemble Kalman method was used to assimilate prior data for quantifying model form uncertainty in RANS models. In Wang \textit{et al.}\cite{wang2017physics,wang2017comprehensive} and Wu \textit{et al.}\cite{wu2018data}, random-forest regressors were utilized for RANS turbulence-modeling given direct numerical simulation (DNS) data. In Singh and Duraisamy \cite{singh2016using} and Singh \textit{et al.}\cite{singh2017machine}, an ANN was utilized to predict a non-dimensional correction factor in the Spalart-Allmaras turbulence model through a field-inversion process using experimental data.  Bypassing functional formulations of a turbulence model (a focus of this study) was also studied from the RANS point of view by Tracey \textit{et al.} \cite{tracey2015machine}. Ling and Templeton \cite{ling2015evaluation} utilized support vector machines, decision trees and random forest regressors for identifying regions of high RANS uncertainty. A deep-learning framework where Reynolds-stresses would be predicted in an invariant subspace was developed by Ling \textit{et al.} \cite{ling2016reynolds}. Machine learning of invariance properties has also been discussed in the context of turbulence modeling \cite{ling2016machine}. The reader is directed to a recent review by Duraisamy \textit{et al.}\cite{duraisamy2018turbulence}, for an excellent review of turbulence modeling using data-driven ideas.

As shown above, the use of data-driven ideologies and in particular artificial neural networks (ANNs) has generated significant interest in the turbulence modeling community for addressing long-standing challenges (also see \cite{sotgiu2018turbulent,zhang2018machine,zhu2019machine,zhang2019application,raissi2019deep} for recent progress). One motivation for the popularity of ANNs is that a multilayered ANN may be optimally trained to universally approximate any non-linear function \cite{hornik1989multilayer}. In addition, the deployment of ANNs is amenable to integration within existing computational frameworks. Greater accessibility to data and ever-improving computing capabilities has also motivated the development of advanced ANN architectures for large-scale learning of complicated physical phenomena such as turbulence. Within the context of LES (and associated with the scope of this paper) there are several investigations into sub-grid modeling using data-driven techniques. In one of the first studies of the feasibility of using learning from DNS based high-fidelity data, Sarghini \textit{et al.}\cite{sarghini2003neural} utilized ANNs for estimating Smagorinsky model-form coefficients within a mixed sub-grid model for a turbulent channel flow. ANNs were also used for wall-modeling by Milano and Koumotsakos \cite{milano2002neural} where it was used to reconstruct the near wall field and compared to standard proper-orthogonal-decomposition techniques. An alternative to ANNs for sub-grid predictions was proposed by King \textit{et al.}\cite{king2016autonomic} where \emph{a priori} optimization was utilized to minimize the $L^2$-error between true and modeled sub-grid quantities in a least-squares sense using a parameter-free Volterra series. Maulik and San \cite{maulik2017neural} utilized an extreme-learning-machine (a variant of a single-layered ANN) to obtain maps between low-pass spatially filtered and deconvolved variables in an \emph{a priori} sense. This had implications for the use of ANNs for turbulence modeling without model-form specification. A more in-depth investigation was recently undertaken by Fukami \textit{et al.}\cite{fukami2018super} where convolutional ANNs were utilized for reconstructing from downsampled snapshots of turbulence. Maulik \textit{et al.} \cite{maulik2018deconvolution} also deployed a data-driven convolutional and deconvolutional operation to obtain closure terms for two-dimensional turbulence. Gamahara and Hattori \cite{gamahara2017searching}, utilized ANNs for identifying correlations with grid-resolved quantities for an indirect method of model-form identification in turbulent channel flow. The study by Vollant \textit{et al.} \cite{vollant2017subgrid} utilized ANNs in conjuction with optimal estimator theory to obtain functional forms for sub-grid stresses. In Beck \textit{et al.}\cite{beck2018neural}, a variety of neural network architectures such as convolutional and recurrent neural networks are studied for predicting closure terms for decaying homogeneous isotropic turbulence. A least-squares based truncation is specified for stable deployments of their model-free closures. Model-free turbulence closures are also specified by Maulik \textit{et al.}\cite{maulik2018deconvolution,maulik2019subgrid} and Wang \textit{et al.}\cite{wang2018investigations}, where sub-grid scale stresses are learned directly from DNS data and deployed in \emph{a posteriori} assessments. King \textit{et al.}\cite{king2018deep} studied generative-adversarial networks and the LAT-NET \cite{hennigh2017lat} for \emph{a priori} recovery of statistics such as the intermittency of turbulent fluctuations and spectral scaling. A detailed discussion of the potential benefits and challenges of deep learning for turbulence (and fluid dynamics in general) may be found in the article by Kutz \cite{kutz2017deep}.

While a large majority of the LES-based frameworks presented above utilize a least-squares error minimization technique for constructing maps to sub-grid stresses \emph{directly} for theoretically optimal LES \cite{langford1999optimal,moser2009theoretically,labryer2015framework}, this work is novel in that it utilizes sub-grid statistics (pre-computed from DNS data) to train a classifier. This classifier determines whether a location requires dissipation or not through \emph{a priori} experience in the learning phase. Once classified, the non-linear term at this particular point is evaluated using one of two schemes. If it is determined that the point requires no sub-grid closure, a symmetric and second-order accurate, energy and enstrophy conserving Arakawa-scheme \cite{arakawa1981potential} is utilized for the non-linear term computation. If dissipation is necessary, an upwinding scheme is utilized instead. Therefore this study may be interpreted as a machine learning framework for devising hybrid schemes for non-linear term computation with a view to reconstructing turbulence statistics in a superior fashion. Therefore, this study is similar to that employed by Ling and Kurzawski \cite{ling2017data} for adaptively determining RANS corrections. We note that the classification framework devised in this study is also deployed in an aligned work to switch between functional and structural explicit LES hypotheses spatio-temporally \cite{maulik2018online} thus proving that high-fidelity DNS statistics may be qualitatively utilized to inform modeling strategies through conditional probability predictions. The article shall describe how the proposed framework is effective in moderating the larger dissipation of an upwinded-scheme through assessments on the Kraichnan turbulence test-case. 

\section{Turbulence modeling equations}

The governing equations for two-dimensional turbulence are given by the Navier-Stokes equations in the vorticity-stream function formulation. In this formulation, our non-dimensional governing equation for incompressible flow may be represented as
\begin{align}
\label{eq1}
\frac{\partial \omega}{\partial t} + J(\omega,\psi) = \frac{1}{Re} \nabla^2 \omega,
\end{align}
where $Re$ is the Reynolds number, $\omega$ and $\psi$ are the vorticity and stream function respectively connected to each other through the Poisson equation given by
\begin{align}
\label{eq2}
\nabla^2 \psi = - \omega.
\end{align}
It may be noted that the Poisson equation implicitly ensures a divergence-free flow evolution. The non-linear term (denoted the Jacobian) is given by
\begin{align}
\label{eq3}
J(\omega,\psi) = \frac{\partial \psi}{\partial y} \frac{\partial \omega}{\partial x} - \frac{\partial \psi}{\partial x} \frac{\partial \omega}{\partial y}.
\end{align}
The stream function and the two-dimensional velocity components are related as 
\begin{align}
\label{eq3a}
u &= \frac{\partial \psi}{\partial y}, \quad v = -\frac{\partial \psi}{\partial x}.
\end{align}

A reduced-order implementation of the aforementioned governing laws (i.e., an LES) is obtained through
\begin{align}
\label{eq4}
\frac{\partial \bar{\omega}}{\partial t} + J(\bar{\omega},\bar{\psi}) = \frac{1}{Re} \nabla^2 \bar{\omega},
\end{align}
where the overbarred variables are now evolved on a grid with far fewer degrees of freedom. Due to the reduction in supported frequencies, the non-linear Jacobian fails to capture inter-eddy interactions at different wavenumbers. If it is assumed that the finer scales of vorticity are generally dissipative in nature for two-dimensional turbulence (based on Kraichnan's cascade of enstrophy \cite{kraichnan1967inertial}), dissipative models may be embedded into the coarse-grained evolution of the vorticity evolution equation to recover some portion of the effect of the finer scales. Explicit LES closures embed dissipation into the vorticity evolution in the form of eddy-viscosity phenomenology or through structural arguments of scale-similarity. However ILES manipulates the computation of the non-linear Jacobian term to add numerical dissipation to mimic that of the unresolved frequencies. The latter framework, while numerically robust, suffers from difficulties associated with \emph{directed} dissipation where it is often very easy to be over-dissipative in regions where sub-grid dissipation may not be as pronounced. In this article, we introduce a hybrid ILES framework that focuses upwinding at areas where high probability of sub-grid dissipation necessity is detected.

\section{Non-linear Jacobian computation}

The study utilizes two types of non-linear term computation schemes. Our first choice is symmetric, second-order accurate and conserves energy and enstrophy to minimize numerical dissipation. This is given by the well-known second-order Arakawa scheme \cite{arakawa1981potential} as detailed below. The non-linear term in Equation \ref{eq4} may be numerically calculated on a coarse grid using 
\begin{align}
J^A (\bar{\omega},\bar{\psi}) = \frac{1}{3} \left( J_1 (\bar{\omega}, \bar{\psi}) + J_2 (\bar{\omega}, \bar{\psi}) + J_3 (\bar{\omega}, \bar{\psi}) \right)
\end{align}
where $J^A(\bar{\omega},\bar{\psi})$ will henceforth refer to the Arakawa discretization. The individual terms on the right hand side of the above equation are given as 
\begin{align}
\begin{split}
 J_1 (\bar{\omega},\bar{\psi}) & = \frac{1}{4 \Delta x \Delta y} \left[ (\bar{\omega}_{i+1,j}-\bar{\omega}_{i-1,j}) (\bar{\psi}_{i,j+1} - \bar{\psi}_{i,j-1}) \right. \\ 
& \left.  - (\bar{\omega}_{i,j+1}-\bar{\omega}_{i,j-1}) (\bar{\psi}_{i+1,j} - \bar{\psi}_{i-1,j}) \right],
\end{split}
\end{align}

\begin{align}
\begin{split}
 & J_2 (\bar{\omega},\bar{\psi}) = \frac{1}{4 \Delta x \Delta y} \left[ \bar{\omega}_{i+1,j} (\bar{\psi}_{i+1,j+1}-\bar{\psi}_{i+1,j-1}) \right. \\ 
 & \left. - \bar{\omega}_{i-1,j} (\bar{\psi}_{i-1,j+1}-\bar{\psi}_{i-1,j-1})  - \bar{\omega}_{i,j+1} (\bar{\psi}_{i+1,j+1}-\bar{\psi}_{i-1,j+1})\right. \\
 & \left. + \bar{\omega}_{i,j-1} (\bar{\psi}_{i+1,j-1}-\bar{\psi}_{i-1,j-1}) \right],
\end{split}
\end{align}

\begin{align}
\begin{split}
& J_3 (\bar{\omega},\bar{\psi}) = \frac{1}{4 \Delta x \Delta y} \left[ \bar{\omega}_{i+1,j+1} (\bar{\psi}_{i,j+1} - \bar{\psi}_{i+1,j}) \right. \\
& \left. - \bar{\omega}_{i-1,j-1} (\bar{\psi}_{i-1,j}-\bar{\psi}_{i,j-1}) - \bar{\omega}_{i-1,j+1} (\bar{\psi}_{i,j+1}-\bar{\psi}_{i-1,j}) \right. \\
& \left. + \bar{\omega}_{i+1,j-1} (\bar{\psi}_{i+1,j}-\bar{\psi}_{i,j-1}) \right].
\end{split}
\end{align}
The aforementioned scheme is utilized when our proposed classifier recognizes that no dissipation is necessary. 

A numerically dissipative computation of the non-linear term allows for that stabilization of noise accumulation at the grid cut-off wavenumbers. Although there are many different methodologies for upwind based dissipation with varying degrees of complexity, in this article, we utilize a conventional upwind-biased scheme as detailed in the following \cite{hoffmann2000computational}. Our ILES Jacobian is computed as 
\begin{align}
\begin{split}
J^I(\bar{\omega},\bar{\psi}) =& \bar{u}_{i,j} \frac{\bar{\omega}_{i+1,j} - \bar{\omega}_{i-1,j}}{2 \Delta x} + \frac{1}{2} (\bar{u}^{+} \bar{\omega}_x^{-} + u^{-} \bar{\omega}_x^{+}) \\
& + \bar{v}_{i,j} \frac{\bar{\omega}_{i,j+1} - \bar{\omega}_{i,j-1}}{2 \Delta y} + \frac{1}{2} (\bar{v}^{+} \bar{\omega}_y^{-} + \bar{v}^{-} \bar{\omega}_y^{+}),
\end{split}
\end{align}
where 
\begin{alignat}{2}
\bar{u}^{-} &= \min(\bar{u}_{i,j},0), \quad \bar{u}^{+} &= \max(\bar{u}_{i,j},0), \\
\bar{v}^{-} &= \min(\bar{v}_{i,j},0), \quad \bar{v}^{+} &= \max(\bar{v}_{i,j},0).
\end{alignat}
In addition,
\begin{align}
\begin{split}
\bar{\omega}_x^{-} &= \frac{\bar{\omega}_{i-2,j} - 3 \bar{\omega}_{i-1,j} + 3 \bar{\omega}_{i,j} - \bar{\omega}_{i+1,j} }{3 \Delta x}, \\
\bar{\omega}_x^{+} &= \frac{\bar{\omega}_{i-1,j} - 3 \bar{\omega}_{i,j} + 3 \bar{\omega}_{i+1,j} - \bar{\omega}_{i+2,j} }{3 \Delta x}, \\
\bar{\omega}_y^{-} &= \frac{\bar{\omega}_{i,j-2} - 3 \bar{\omega}_{i,j-1} + 3 \bar{\omega}_{i,j} - \bar{\omega}_{i,j+1} }{3 \Delta y}, \\
\bar{\omega}_y^{+} &= \frac{\bar{\omega}_{i,j-1} - 3 \bar{\omega}_{i,j} + 3 \bar{\omega}_{i,j+1} - \bar{\omega}_{i,j+2} }{3 \Delta y}.
\end{split}
\end{align}
Note that velocity components are recovered using
\begin{align}
\begin{split}
\bar{u}_{i,j} &= \frac{\bar{\psi}_{i,j+1}-\bar{\psi}_{i,j-1}}{2 \Delta y} \\
\bar{v}_{i,j} &= -\frac{\bar{\psi}_{i+1,j}-\bar{\psi}_{i-1,j}}{2 \Delta x},
\end{split}
\end{align}
where the second-order accurate reconstruction of the velocity leads to overall second-order accuracy for non-linear Jacobian reconstruction using the upwinded procedure outlined above. We also note that our Poisson equation given by Equation \ref{eq2} is solved using a spectrally-accurate scheme. 

With the choice of one of the two aforementioned schemes, a point in space-time may or may not have an artificial dissipation imparted to it numerically. However, we mention the caveat that switching between these two schemes would mean that the kinetic energy and enstrophy preserving property of the Arakawa scheme is lost. 

\section{Machine learning for scheme selection}

\begin{figure*}
\centering
\includegraphics[width=\textwidth]{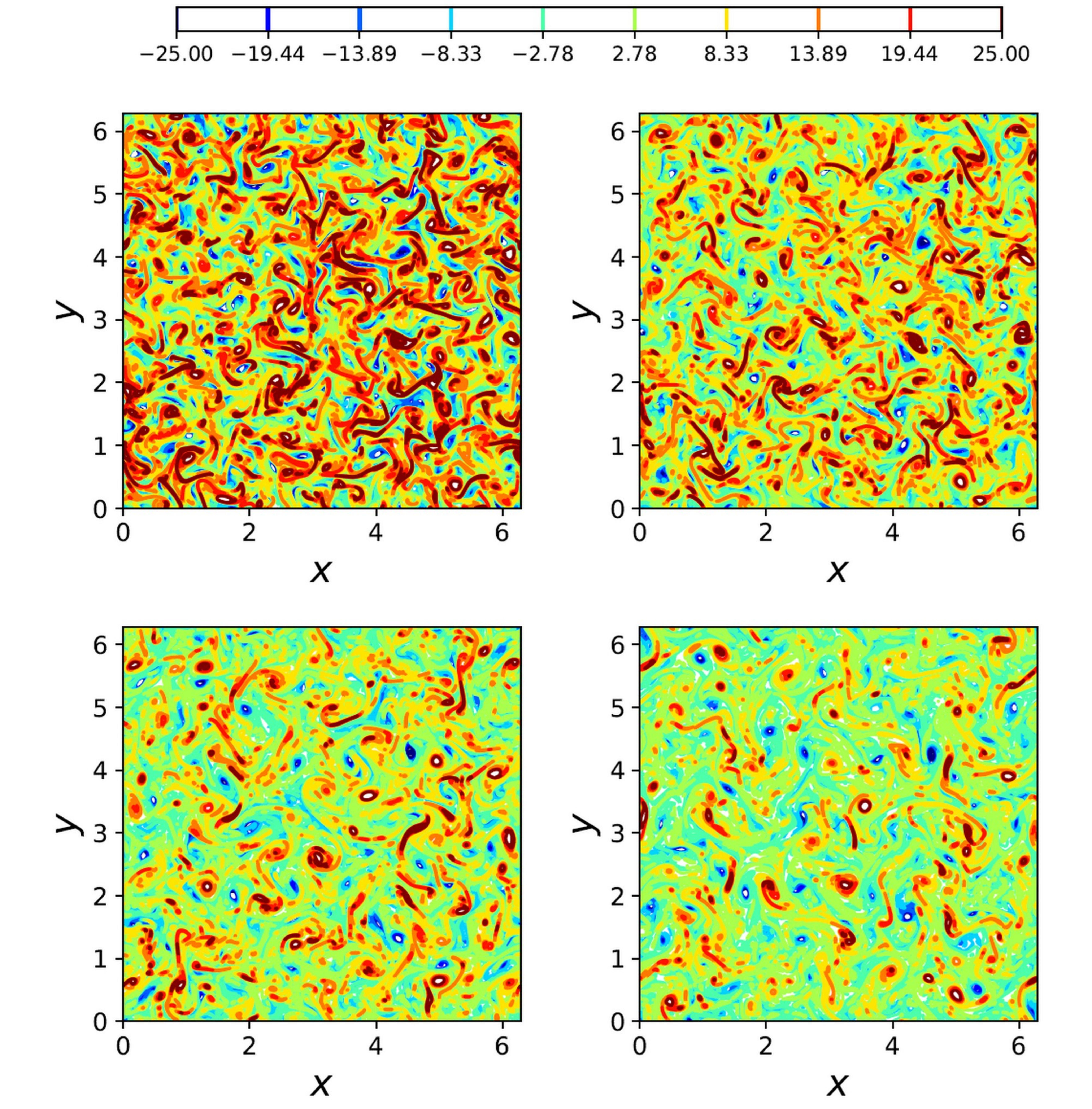}
\caption{Time evolution of the Kraichnan turbulence case with DNS ($N^2 = 2048^2$) contours for vorticity of $t=1$ (top-left), $t=2$ (top-right), $t=3$ (bottom-left), $t=4$ (bottom-right). One can discern the dissipation of vorticity as the system evolves.}
\label{Fig1}
\end{figure*}

We now discuss the procedure of utilizing DNS data for learning to classify one of the two dissipation scenarios. Of these two options, one is given by the choice of the Arakawa scheme and the other by our upwinded computation of the Jacobian (i.e., when the classification framework has determined that the point does not require sub-grid dissipation or vice-versa respectively). This switching between scenarios is spatio-temporally dynamic. We proceed by outlining our training strategy through the utilization of DNS data. Five equidistant snapshots of DNS data at $Re=32000$ (i.e., at $t=0,1,2,3,4$) and at $N^2 = 2048^2$ degrees of freedom (from 40000 available snapshots) are utilized to compute the grid-filtered variables (denoted FDNS) (at $N^2 = 256^2$ degrees of freedom) through the application of a spectral cut-off filter. Perfect closure values 
\begin{align}
\Pi = J(\bar{\omega},\bar{\psi})-\overline{J(\omega,\psi)}
\end{align}
are then obtained (the reader is directed to \citep{maulik2019subgrid} for details related to the calculation of these quantities). Note here, that the Kraichnan turbulence problem is transient with the evolution of vorticity represented in Figure \ref{Fig1} representing different closure needs over time evolution. 

We proceed by introducing the \emph{a priori} eddy-viscosity given by
\begin{align}
\nu_e^a = \frac{\Pi}{\nabla^2 \bar{\omega}}
\end{align}
where the right-hand side of the above equation may be calculated from DNS snapshots. The \emph{a priori} eddy-viscosity is centered at zero (corresponding to where closure modeling is unnecessary) and spreads out in the negative and positive directions (a hallmark of isotropic turbulence). We segregate this \emph{a priori} estimate of sub-grid effects into three categories as follows. The \emph{a priori} eddy-viscosities calculated from the DNS data are compared with a Gaussian distribution where values lying less than a distance of 1\% of the standard-deviation from the mean (which is zero) are labeled as those requiring no dissipation (due to the low strength of the \emph{a priori} eddy-viscosity). For posterity, we label these points as $k=1$. Positive values lying beyond this range are labeled as those requiring sub-grid dissipation and are labeled $k=2$. Negative values less than 1\% of the standard-deviation are also considered to require no dissipation and are labeled $k=3$. This three-category segregation stems from a learning hypothesis that seeks to identify regions in a flow evolution that require structural, functional or no-closure modeling hypothesis. We link labels of negative or nearly-zero eddy-viscosities to the Arakawa classification and positive eddy-viscosities to the upwinded classification. The positive eddy-viscosity prediction would indicate that the sub-grid term at a point is predominantly dissipative in nature at which point the numerical dissipation of the upwinded scheme would be utilized. We note here that the concept of an \emph{a priori} eddy-viscosity lies firmly within the explicit LES hypothesis. The classifier is therefore instrumental in moderating ILES deployments through a decision making process that recognizes the dissipative (or forcing) nature of the sub-grid quantities. 

We note that the choice of 1\% as the decision parameter for switching between hypothesis is motivated by a sensitivity study that showed the highest classification accuracy for the ANN framework. Larger choices of this hyper-parameter would result in a classifier that would be prone to classify most points in the `no-model' zone. However, we clarify that the choice of this value is also correlated with the architecture of the ANN. A potential extension of the proposed hypothesis is to combine architecture search algorithms with varying value of the decision hyper-parameters for larger classification accuracies. In addition, the three-category framework is derived from an aligned study \cite{maulik2018online} where sub-grid models are determined according to negative, positive and nearly-zero \emph{a priori} eddy-viscosities and utilizes the same learning. This enables use to determine a unified framework for switching between turbulence model hypotheses as well as numerical dissipation scenarios. However, we would like to emphasize that, for the purpose of switching between the Arakawa and upwinded Jacobian computation, a simple two-class framework would also suffice. 

\begin{figure}
\centering
\includegraphics[trim={3cm 16cm 6cm 1cm},clip,width=\columnwidth]{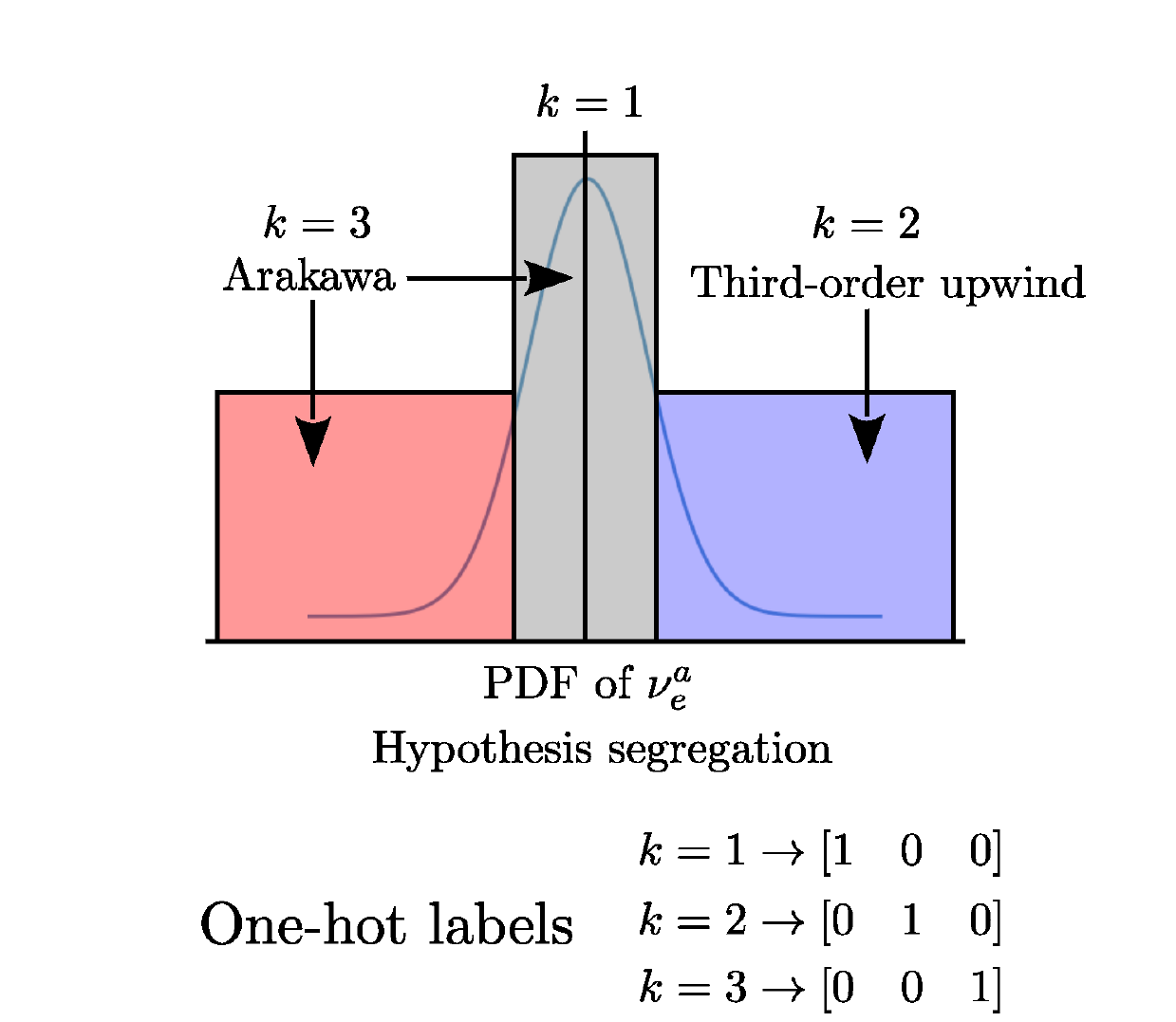}
\caption{Hypothesis segregation and one-hot labeling for our proposed framework. The learning predicts conditional probabilities for the three segregated \emph{a priori} eddy-viscosity classes which are utilized for Jacobian calculation decisions spatio-temporally.}
\label{Segregation}
\end{figure}

\begin{figure}
\centering
\includegraphics[width=\columnwidth]{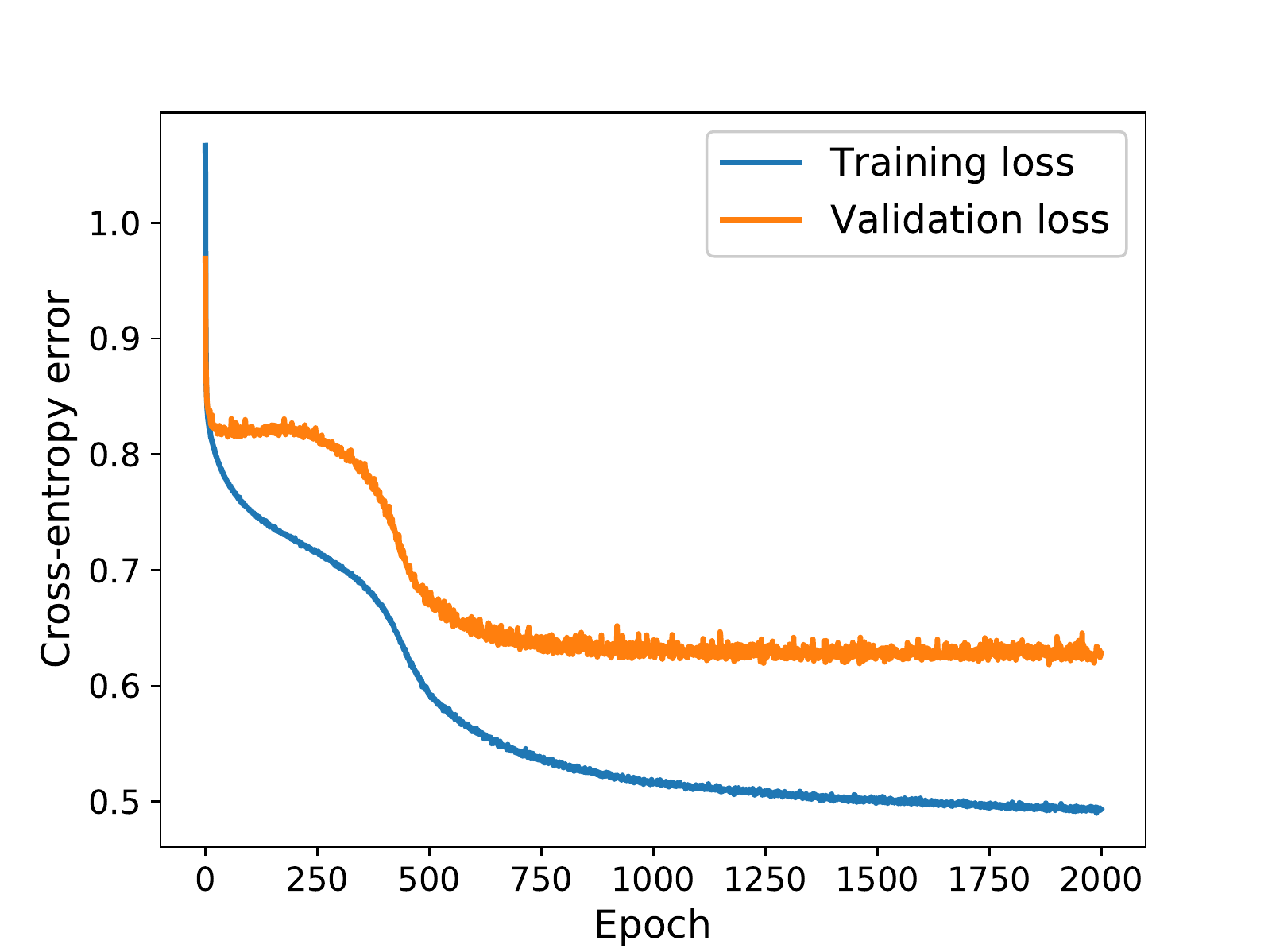}
\caption{Learning rate and convergence of our classification framework training. 2000 epochs were sufficient for converged validation loss.}
\label{Loss_history}
\end{figure}

A one-hot labeling of our eddy-viscosity classes is utilized for a classification deployment and a schematic for this hypothesis segregation and labeling is shown in Figure \ref{Segregation}. The labels indicate the conditional probability of a point belonging to each possible class. As such, the training labels are given by a value of 1 for the particular class that a point belongs to and zeros for other choices. This is because there is no ambiguity in the class a training sample belongs to. Each label for the \emph{a priori} eddy-viscosity is also associated with a corresponding input kernel of grid-resolved quantities. This kernel is given by a local stencil of vorticity and stream function. There are 9 inputs each for vorticity and stream function given by a query of the field quantity at a point on the coarse grid, 4 adjacent points in each dimension ($x,y$) and the 4 diagonally adjacent points. Each sample of our training data thus consists of 18 inputs of vorticity and stream function and outputs given by one-hot labels for the choice of closure modeling strategy. We then utilize an ANN to establish a relationship between these inputs and outputs. Mathematically, if our input vector $\mathcal{P}$ resides in a $P$-dimensional space and our desired output $\mathcal{Q}$ resides in a $Q$-dimensional space, this framework establishes a map $\mathbb{M}$ as follows:
\begin{align}
\label{eq6}
\mathbb{M} : \{ \mathcal{P}_1, \mathcal{P}_2, \hdots, \mathcal{P}_P\} \in \mathbb{R}^P \rightarrow \{ \mathcal{Q}_1, \mathcal{Q}_2, \hdots, \mathcal{Q}_Q\} \in \mathbb{R}^Q.
\end{align}
Accordingly, the framework utilized in this article leads to the following relation:
\begin{align}
\label{eq7}
\mathbb{M} : \{ \textbf{p} \} \in \mathbb{R}^{18}  \rightarrow \{ P(\textbf{q}|\textbf{p})\} \in \mathbb{R}^3,
\end{align}
where 
\begin{align}
\begin{gathered}
\textbf{p}_{i,j} = \{ \bar{\omega}_{i,j}, \bar{\omega}_{i,j+1}, \bar{\omega}_{i,j-1}, \hdots, \bar{\omega}_{i-1,j-1}, \\ \bar{\psi}_{i,j}, \bar{\psi}_{i,j+1}, \bar{\psi}_{i,j-1}, \hdots, \bar{\psi}_{i-1,j-1} \}
\end{gathered}
\end{align}
is our input vector for each query of the machine learning framework and where 
\begin{align}
P(\textbf{q}|\textbf{p})_{i,j} = \{ P(J^k(\bar{\omega},\bar{\psi})_{i,j}| \textbf{p}_{i,j})\},
\end{align}
is the conditional probability of a Jacobian computation (given by a connection to the explicit closure hypothesis). Note that $i,j$ refer to the spatial indices on the coarse-grid (i.e., the point of deployment). The indices $k=1$ and $k=3$ refer to the Arakawa non-linear Jacobian computation and $k=2$ refers to the upwinded computation instead (see Figure \ref{Segregation}). Our optimal map $\mathbb{M}$ is then trained by minimizing the categorical cross-entropy loss-function
\begin{align}
E(\textbf{w}) = -\sum_{n=1}^{N} \sum_{k=1}^{K} \{ t_{nk} \log(y_{nk}) + (1-t_{nk})\log(1-y_{nk})\},
\end{align}
where $\textbf{w}$ are the variable weight and bias parameters of the network, $N$ refers to the total number of samples and $K=3$ is the total number of classification scenarios (i.e., negative, positive or nearly-zero \emph{a priori} eddy-viscosities). Here, $t_{nk}$ refers to the true label of class $k$ and sample $n$ and $y_{nk}$ refers to a corresponding prediction of the learning framework. One-hot encoding ensures that $t_{nk}$ values are always binary \cite{Bishop:2006:PRM:1162264} and the outputs of the ANN may be interpreted as conditional-probabilities. Our optimal architecture is given by five 40-neuron hidden layers (obtained via grid-search hyper-parameter tuning). All hidden layers utilize ReLU units to impart non-linearity to the layer-wise transformations. For reference, our architecture is trained using the open-source deep learning software Tensorflow and is optimized with the use of ADAM, a popular gradient-descent based optimizer \cite{kingma2014adam}. Figure \ref{Loss_history} shows the progress to convergence for our framework with our optimally trained network displaying approximately 79\% accuracy in classifying points to their correct labels. To summarize this section, we train a deep ANN to estimate probabilities of negative, positive or nearly-zero eddy-viscosities which are utilized to decide the choice of the Jacobian computation. We clarify that the decision to deploy a particular hypothesis is obtained by utilization of the classification scenario which has the highest conditional probability.

\section{Results}

\subsection{\emph{A posteriori} deployment}

\begin{figure}
\centering
\includegraphics[width=\columnwidth]{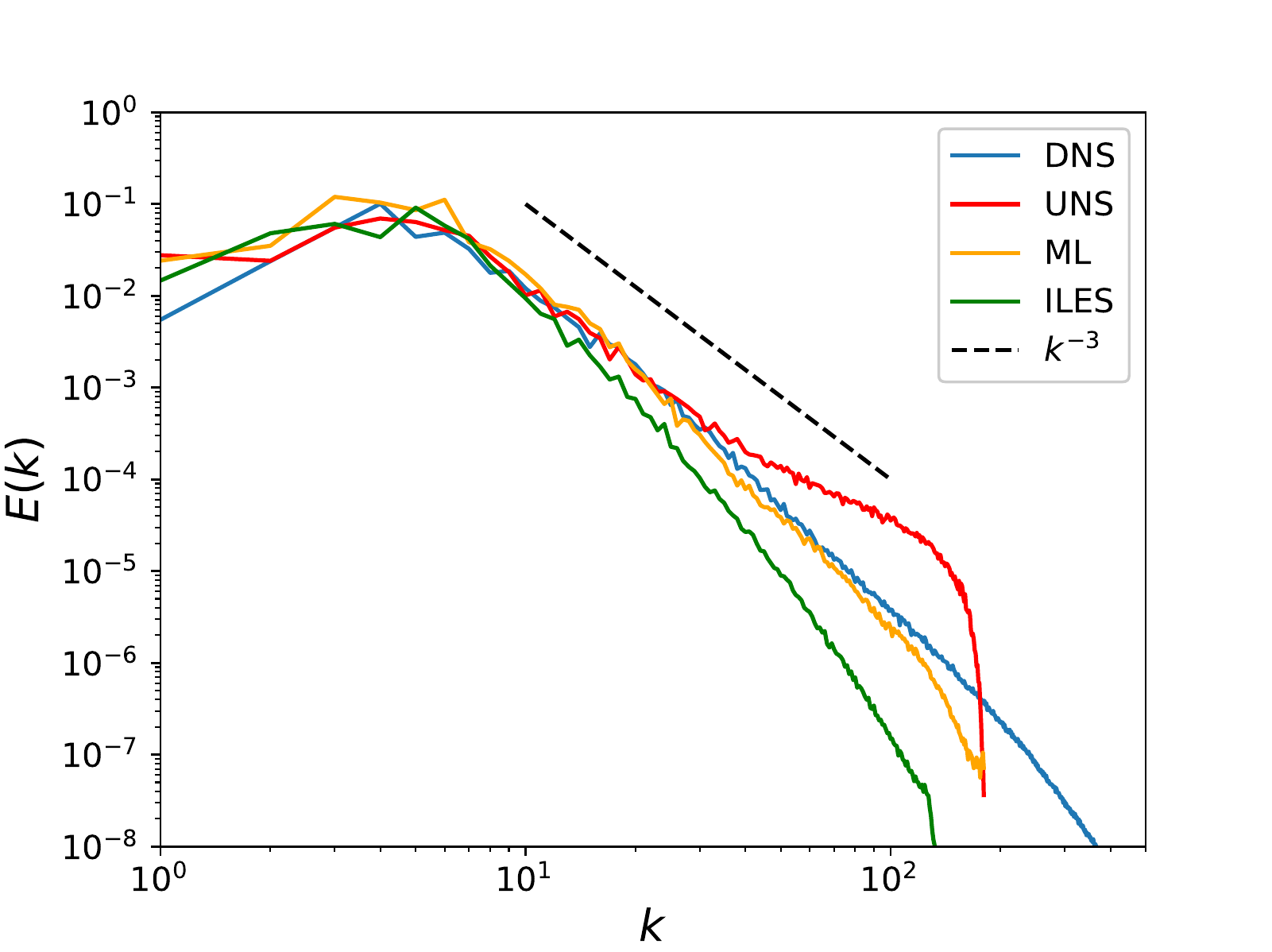}
\caption{The \emph{a posteriori} performance of proposed framework (ML) for $Re=32000$ and at $t=4$ in terms of angle-averaged kinetic energy spectra. Comparisons with DNS, the Arakawa scheme (UNS) and the upwinded scheme (ILES) show that ML provides directed dissipation adequately.}
\label{Fig2}
\end{figure}

In this section, we detail the results from an \emph{a posteriori} deployment of the classification framework (denoted ML henceforth) for the Kraichnan test-case. In the LES evolution of the problem, a considerably coarser grid is used (at $N^2=256^2$). We remark that the forward deployment of our framework needs to overcome the challenge of numerical errors and is a robust test of the generalizability and robustness of our learning. Our LES results are assessed using angle-averaged kinetic energy spectra and through structure functions of vorticity. In addition, qualitative comparisons are also provided through visual examinations of the vorticity contours. We remark that the LES deployment is performed from $t=0$ to $t=4$ which spans the training regime data obtained from DNS. In what follows we note that DNS refers to a high-fidelity evolution of the governing equations (i.e., at $N^2=2048^2$ degrees of freedom), UNS refers to results obtained using the Arakawa scheme alone and ILES refers to a simulation where the non-linear Jacobian at all points in space and time are upwinded. Figure \ref{Fig2} shows the \emph{a posteriori} performance of the proposed framework at $Re=32000$ in terms of energy spectra predictions. The reader may find an exact definition of the kinetic-energy spectra in Maulik and San \cite{maulik2017stable}. We note that the training data was obtained for the same Reynolds number as well. The prediction of the proposed framework is seen to agree remarkably well with DNS. It is apparent that the switching of schemes using the classifier has obtained an optimal balance between both techniques.

\begin{figure*}
\centering
\includegraphics[width=\textwidth]{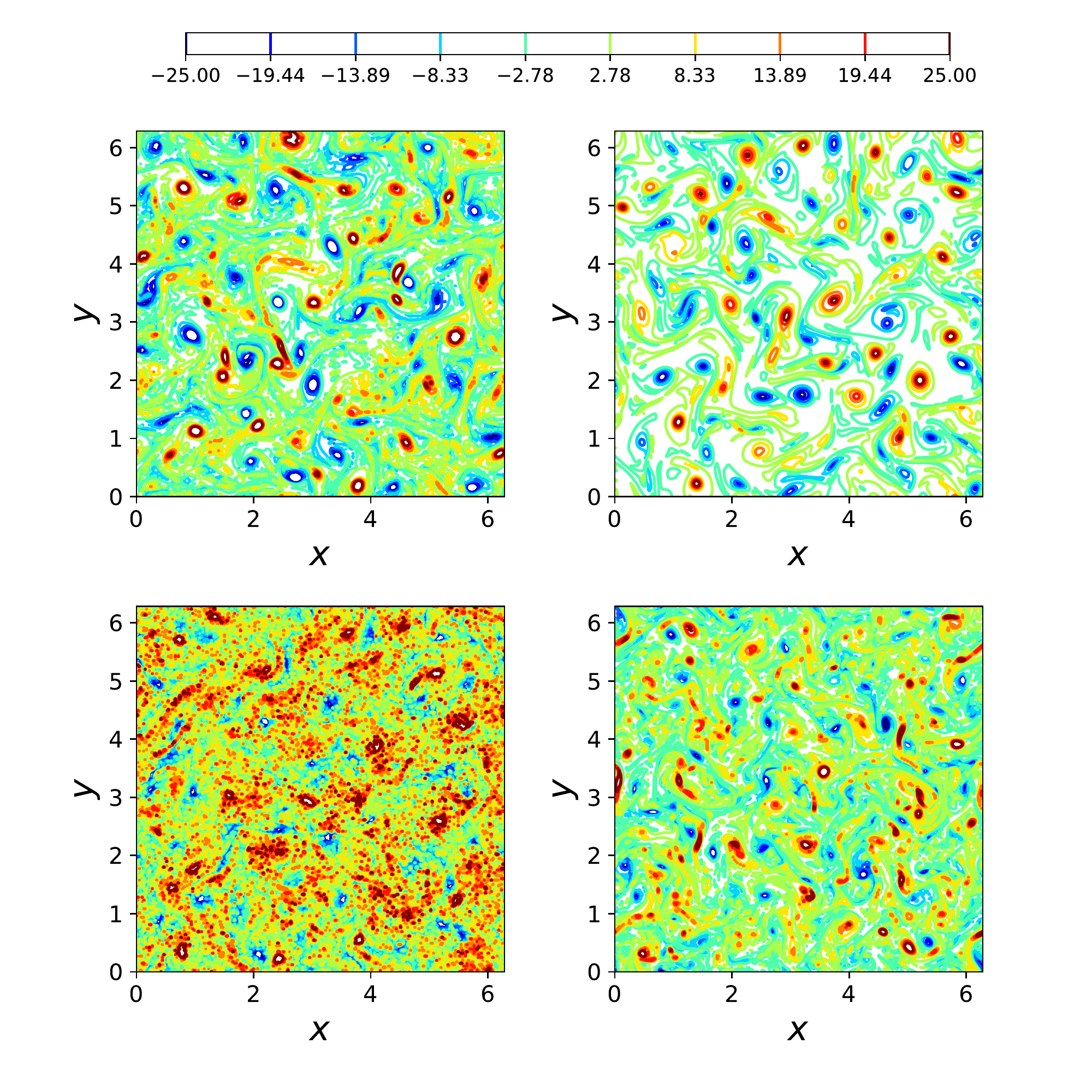}
\caption{Contours for the vorticity at LES resolution and at $t=4$. In the top-left, we have predictions from the ML approach. The top-right field has been obtained using ILES, the bottom-left field is obtained from UNS and the bottom right shows FDNS contours obtained by spectral cut-off filtering of DNS. }
\label{Fig3}
\end{figure*}

Vorticity contours for LES resolution assessments are shown in Figure \ref{Fig3}, where it is apparent that the proposed framework optimally balances the energy-conserving and dissipative natures of the Arakawa and upwinded schemes respectively. This is verified by qualitative examination with FDNS contours obtained by spectrally filtering the DNS snapshot for $Re=32000$ at $t=4$.

\begin{figure}
\centering
\includegraphics[width=\columnwidth]{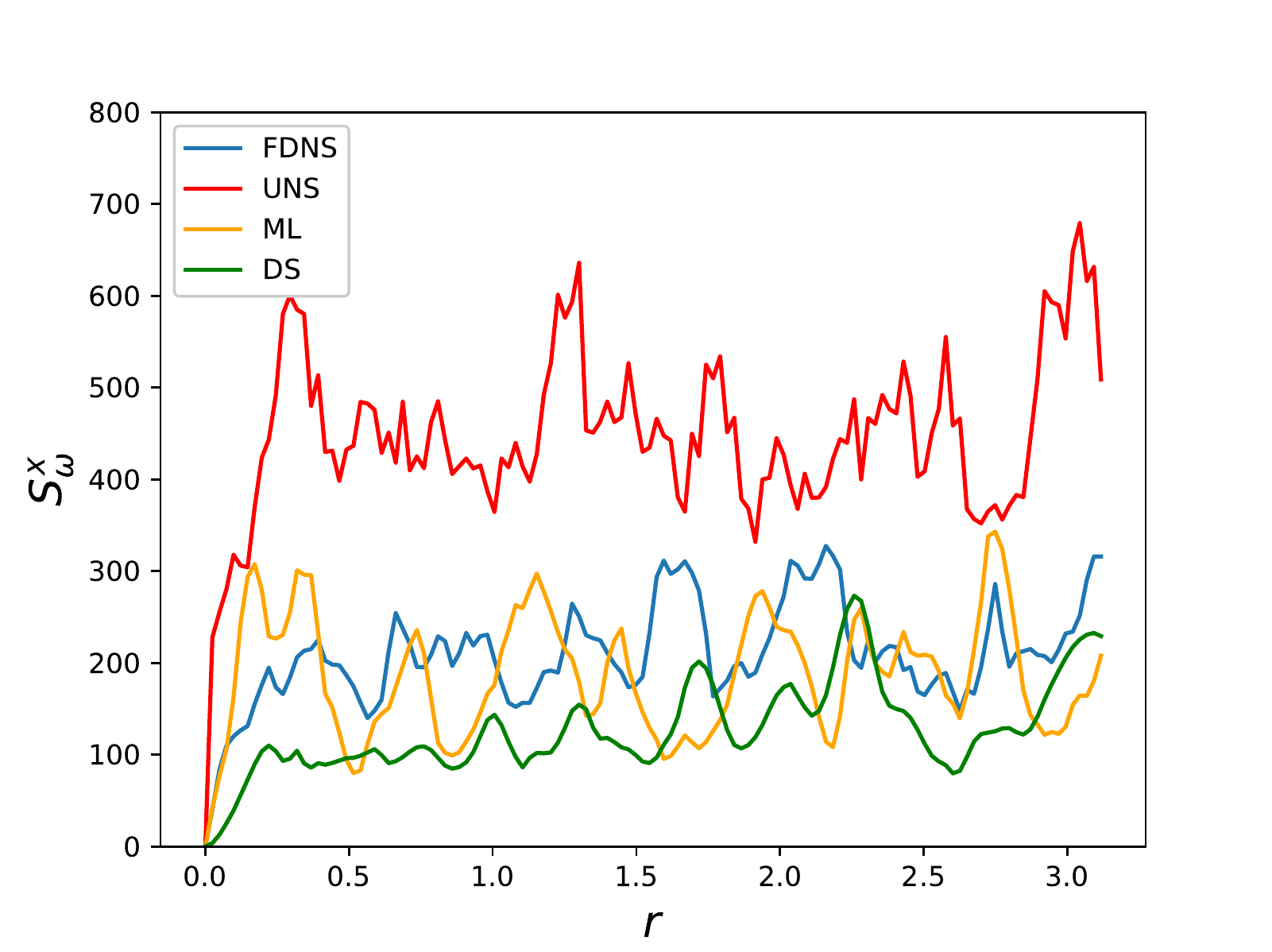}
\caption{\emph{A posteriori} vorticity structure functions in $x$ direction of our proposed framework (ML), the Arakawa scheme (UNS) and the upwind scheme (ILES) with statistics obtained from an FDNS snapshot at $t=4$. It is apparent that the ML method stabilizes the UNS result optimally.}
\label{Fig4}
\end{figure}

\begin{figure}
\centering
\includegraphics[width=\columnwidth]{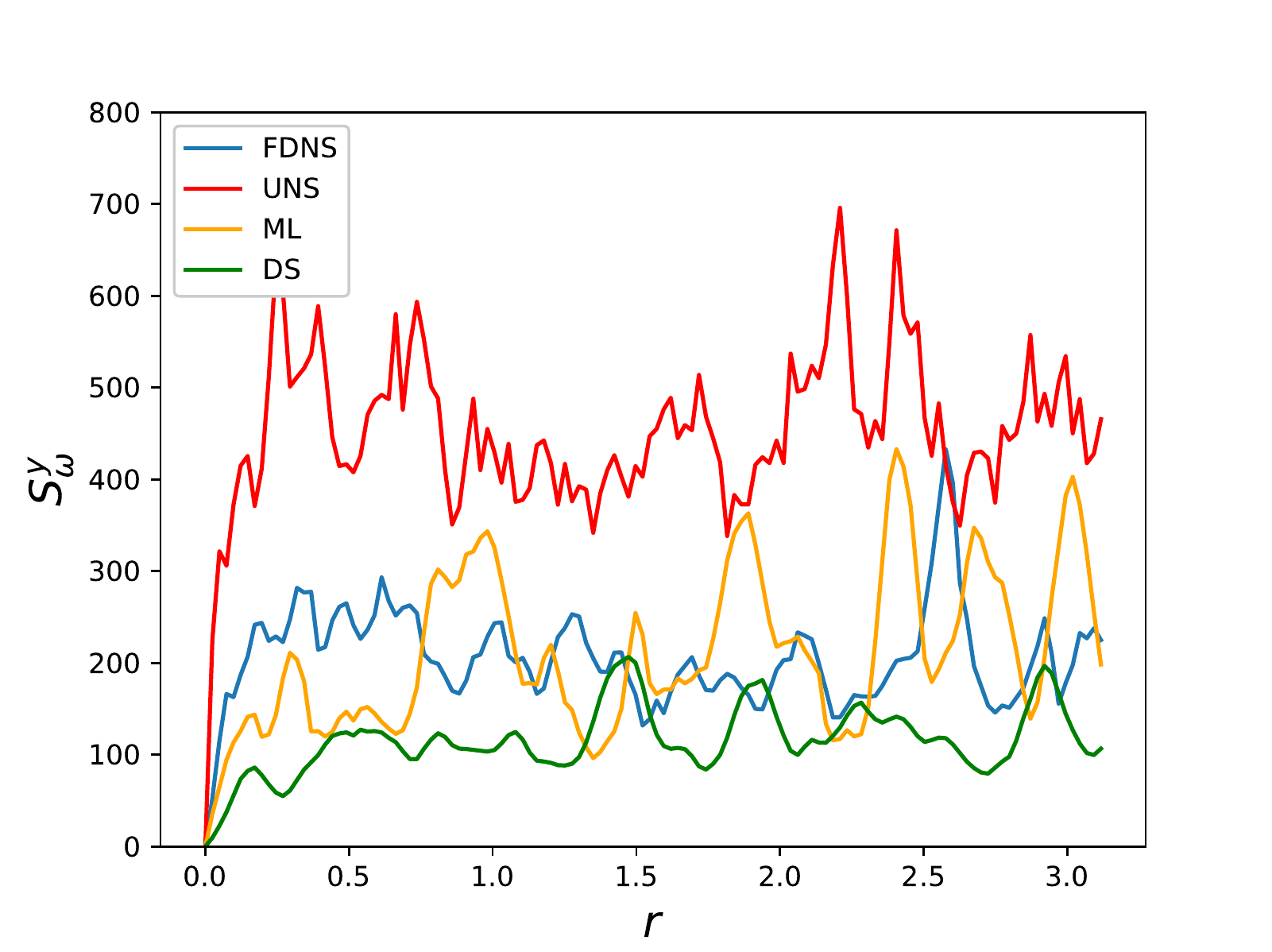}
\caption{\emph{A posteriori} vorticity structure functions in $y$ direction of our proposed framework (ML), the Arakawa scheme (UNS) and the upwind scheme (ILES) with statistics obtained from an FDNS snapshot at $t=4$. It is apparent that the ML method stabilizes the UNS result optimally.}
\label{Fig5}
\end{figure}

A second statistically significant quantity of interest studied in this investigation is the vorticity structure function \cite{grossmann1992structure} given by
\begin{align}
S_\omega^x = \langle |\omega(x+r,y) - \omega(x,y)|^2 \rangle \\
S_\omega^y = \langle |\omega(x,y+r) - \omega(x,y)|^2 \rangle,
\end{align}
where the angle-brackets indicate ensemble averaging and $x,y$ indicate a position on the grid with $r$ being a certain distance from this location. Figures \ref{Fig4} and \ref{Fig5} show the structure functions obtained from \emph{a posteriori} deployments of the UNS, ILES and ML frameworks compared against those obtained from the final time FDNS snapshot. It is clear that the proposed framework balances between UNS and ILES deployments well to recover appropriate trends. We can thus claim that our learning is appropriate for hybrid deployments of dissipative and conservative frameworks for two-dimensional turbulence. Before moving on, we would like to point out to the reader here that the proposed methodology for closure does not require any post-processing prior to deployment in the forward simulation as utilized in several data-driven turbulence modeling studies\cite{beck2018neural,maulik2019subgrid}.

\subsection{Validation of learning}

\begin{figure}
\centering
\includegraphics[width=\columnwidth]{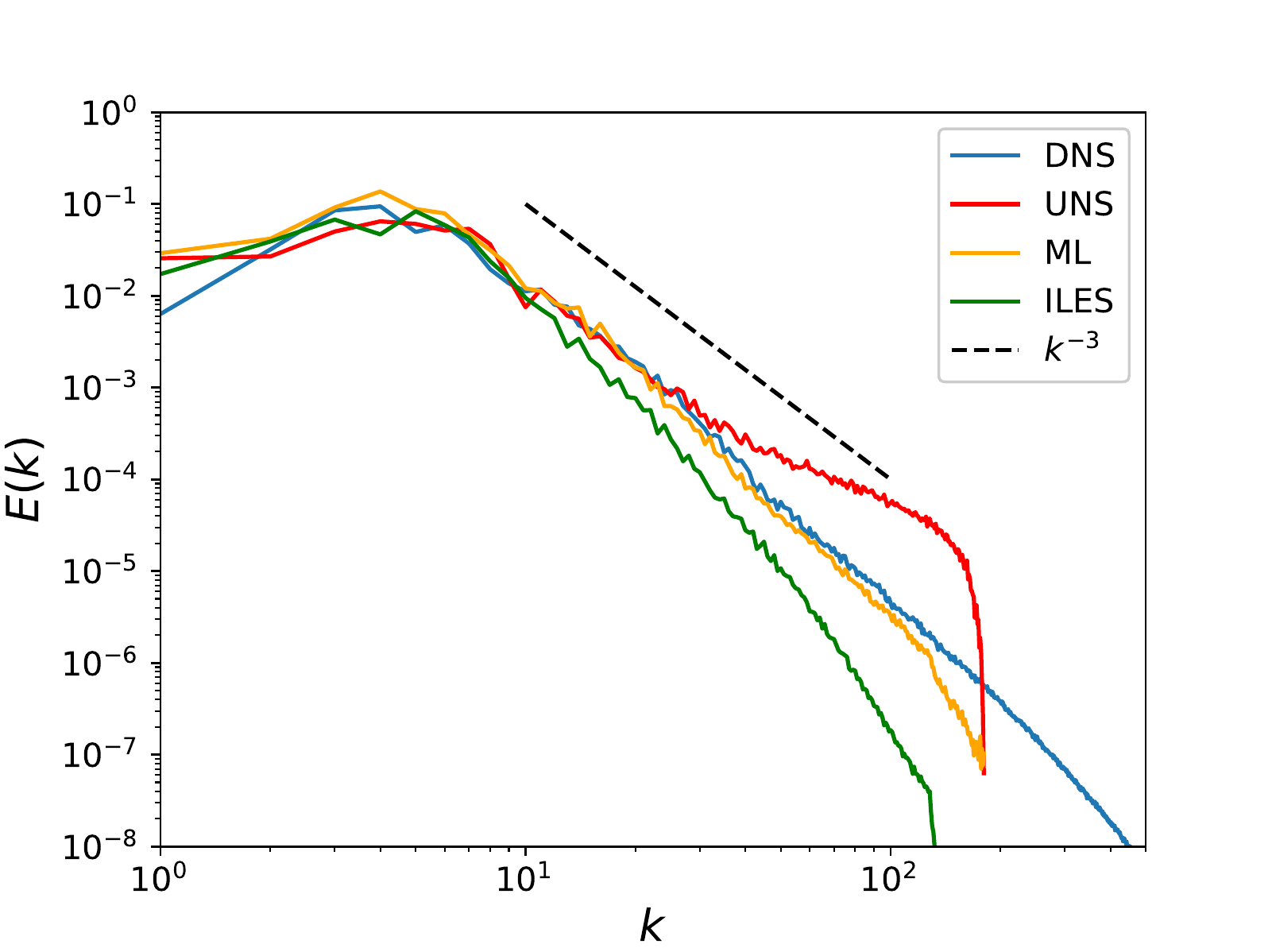}
\caption{The \emph{a posteriori} performance of proposed framework (ML) for $Re=64000$ and at $t=4$ in terms of energy spectra. This represents deployment of our learning at a different Reynolds number than that used for generating training data.}
\label{Fig6}
\end{figure}

\begin{figure}
\centering
\includegraphics[width=\columnwidth]{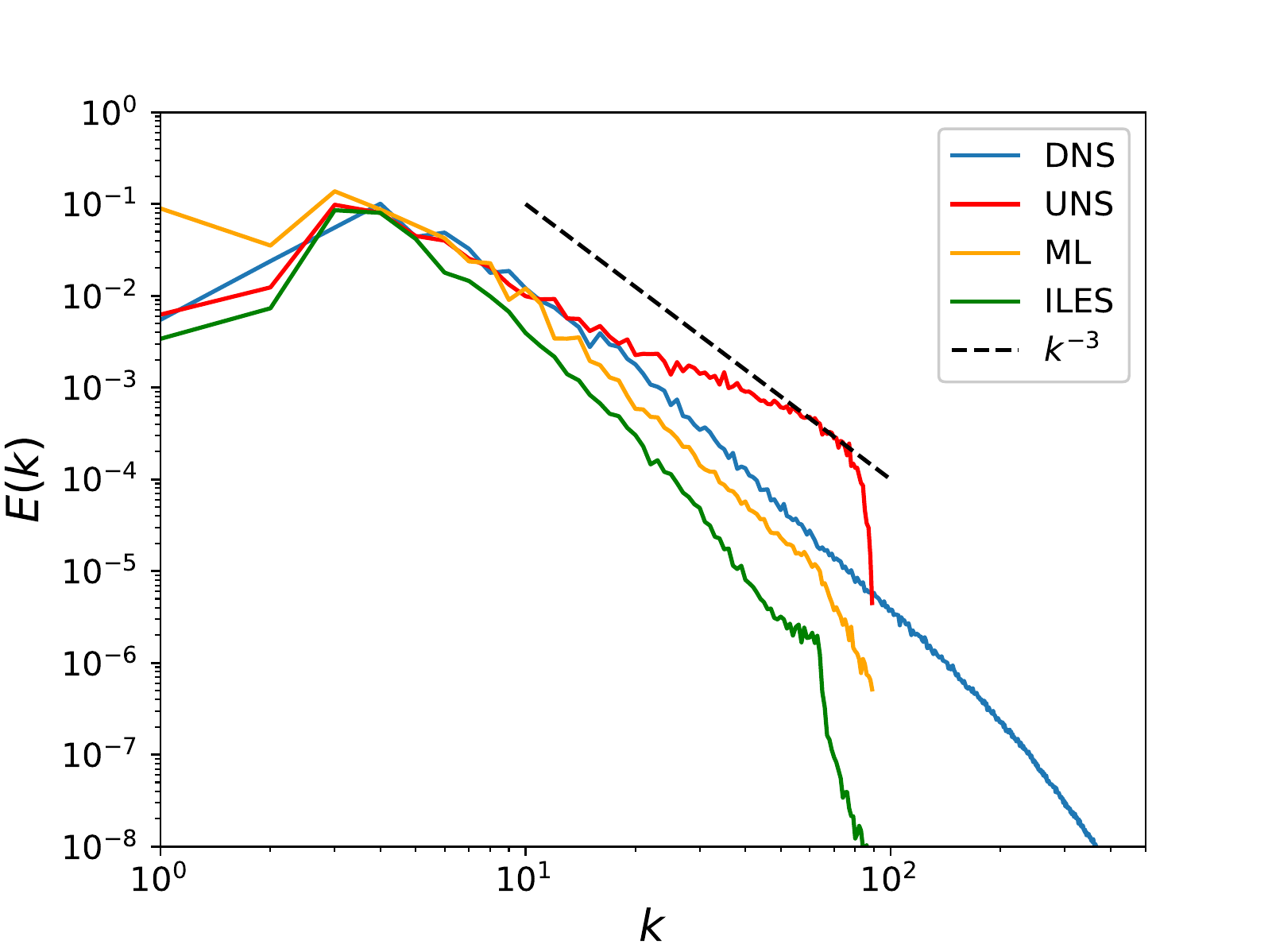}
\caption{The \emph{a posteriori} performance of proposed framework (ML) for $Re=32000$, $t=4$ and at $N^2 = 128^2$ in terms of energy spectra. This represents deployment of our learning at a different resolution than that used for generating training data.}
\label{Fig7}
\end{figure}

In this section, we proceed with a rigorous validation of our learning for deployment in regimes that are not a part of the training data. This is to ensure that the framework has truly learnt a classification based on the underlying physical hypothesis used for data segregation and is not memorizing data. This ensures that our classifier can be used in a more generalizable fashion. Figure \ref{Fig6} shows kinetic energy spectra obtained from the forward deployment of the ML framework for a $Re=64000$ which represents a classification task that the framework has not previously seen (although the physics of the test-case remains similar). As observed, the proposed method performs quite well in this out-of-training data range as well. We note that a similar resolution ($N^2=256^2$) is utilized for this deployment. In contrast, Figure \ref{Fig7} shows the performance of the ML technique for a reduced resolution of $N^2=128^2$ but utilizing the same Reynolds number of 32000. The kinetic energy spectra show a successful stabilization of the flow evolution at this reduced resolution although some forcing to the large scales is observed. This suggests that the classification framework may be improved by sampling from different resolutions. 

\section{Concluding remarks}

In this article, we have proposed a neural network based classifier that enables us to take decisions on the choice of non-linear term computation in the LES evolution of the Kraichnan turbulence test-case. The classifier outputs conditional probabilities for the presence (or absence) of eddy-viscosities within three different ranges during deployment and is used to switch between the Arakawa and upwind computation of the non-linear Jacobian for a hybrid upwinded deployment that optimally directs dissipation on the coarse-grained flow field. Our machine learning framework is trained by calculating \emph{a priori} eddy-viscosities which are projected onto a Gaussian distribution and segregated into three categories. Each category is devised to capture a unique behavior of the underlying sub-grid terms with negative and nearly-zero eddy-viscosity classes signifying absence of sub-grid dissipation. An optimally trained classifier is then utilized to identify if a point requires sub-grid dissipation based on if it is placed in the positive eddy-viscosity category. If so, the upwind Jacobian is calculated for imparting numerical dissipation. 

We perform \emph{a posteriori} assessments on the Kraichnan turbulence test-case through statistical quantities such as the angle-averaged kinetic energy spectra and the vorticity structure functions. It is observed that the proposed framework is successful in balancing the dissipative nature of the upwind scheme and the energy-conserving Arakawa scheme to give excellent agreement with DNS statistics. Validation for out-of-training regimes also indicate that the framework is able to learn the link between grid-resolved quantities at a coarse resolution and the nature of the sub-grid forcing. 

Our conclusions therefore point toward the possibility of using classifiers for the unified deployment of numerical schemes with varying dissipation through the decision making process described above. A key strength of our hypothesis stems from the fact that an ILES deployment is moderated by concepts drawn from the explicit LES ideology (i.e., that of an \emph{a priori} eddy-viscosity). The successful deployment of our method thus points towards the possibility of deploying directed numerical dissipation that preserves the statistics of turbulence without sacrificing the shock-capturing ability of many non-oscillatory schemes. Our future work lies in that particular direction.

\begin{acknowledgements}
This material is based upon work supported by the U.S. Department of Energy, Office of Science, Office of Advanced Scientific Computing Research under Award Number DE-SC0019290. OS gratefully acknowledges their support. Disclaimer: This report was prepared as an account of work sponsored by an agency of the United States Government. Neither the United States Government nor any agency thereof, nor any of their employees, makes any warranty, express or implied, or assumes any legal liability or responsibility for the accuracy, completeness, or usefulness of any information, apparatus, product, or process disclosed, or represents that its use would not infringe privately owned rights. Reference herein to any specific commercial product, process, or service by trade name, trademark, manufacturer, or otherwise does not necessarily constitute or imply its endorsement, recommendation, or favoring by the United States Government or any agency thereof. The views and opinions of authors expressed herein do not necessarily state or reflect those of the United States Government or any agency thereof.
\end{acknowledgements}

\bibliography{references}
\end{document}